# 5 years of survey on the Crab Nebula with SPI/INTEGRAL

**E. Jourdain**
*CESR/UT-CNRS*
*9 avenue du Colonel Roche,31028 Toulouse cedex 04, France*
*E-mail: jourdain@cesr.fr*

**J.P. Roques**
*CESR/UT-CNRS*
*9 avenue du Colonel Roche,31028 Toulouse cedex 04, France*
*E-mail:* `roques@cesr.fr`

We present observations of the Crab Nebula above 20 keV by the SPI/INTEGRAL telescope during more than 5 years of operations. Our study demonstrates the stability of the instrument with time and allows a detailed analysis of the emission observed from the Crab Nebula between 20 keV and 1 MeV. The flux stability is discussed and serves a robust spectral shape analysis. We find that a single power law is clearly excluded since the photon spectrum presents a curvature in the considered energy domain. We have modelled it by a broken power law with the energy break fixed to 100 keV and determined the two photon indices together with the 100 keV flux for 9 periods between 2004 and 2008. The spectral shape of the Crab nebula is very stable as well as its intensity and connects nicely with previous measurements, at lower (X-rays) or higher (MeV) energies.

## 1. Introduction

The SPI telescope aboard the INTEGRAL mission, launched in October 2002, is in orbit since more than 5 years. It is the best operational instrument to study the sky in the hard X-rays (E > 20 keV) domain. We take benefit of its stability to perform a deep analysis of the Crab Nebula, a major source for the high energy astrophysics.

## 2. The SPI instrument

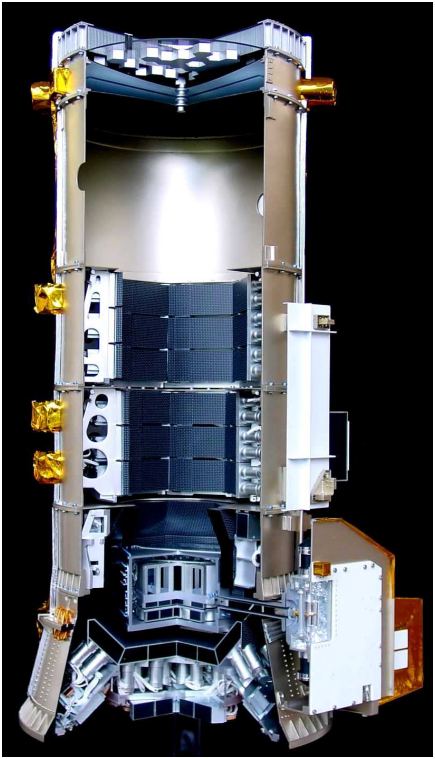

Figure 1: The instrument view
The SPI telescope is based on the association of a coded mask and a gamma camera.. The detector plane consists of 19 hexagonal crystals of high purity germanium operating at 80 K.
The dimensions of an individual detector are 5.6 cm (flat to flat) by 6.94 cm height leading to a total geometrical area of ~ 500 cm2.

The main caracteristics are the followings:

- Imaging : 16° fully coded FOV
- Angular resolution: 2.6°
- Energy range : 20 keV-8 MeV
- Energy resolution: 0.2 %
- Time resolution: 100 μsec
- Shield: active BGO shield
- Camera : 19 HP Ge detectors.
- Active cooling: 80-85 K

More informations on the SPI telescope and its performance can be found in [1, 2].

## 3. The Crab Observations:

INTEGRAL performs regular observations dedicated to the Crab Nebula, twice per year, in february-March and September-October, for calibrations purposes. For SPI, the instrument response matrices are those built from ground calibrations (see [3, 4] for details) and we use these campaigns only to quantify the evolution of the efficiencies and other instrumental caracteristics.

**3.1 Light curve analysis**

In Fig. 2, three observations performed along the mission are compared from 25 to 300 keV. The light curves extracted at the Crab position on the ScW timescale prove that the SPI telescope is remaining very stable after 5 years in space.

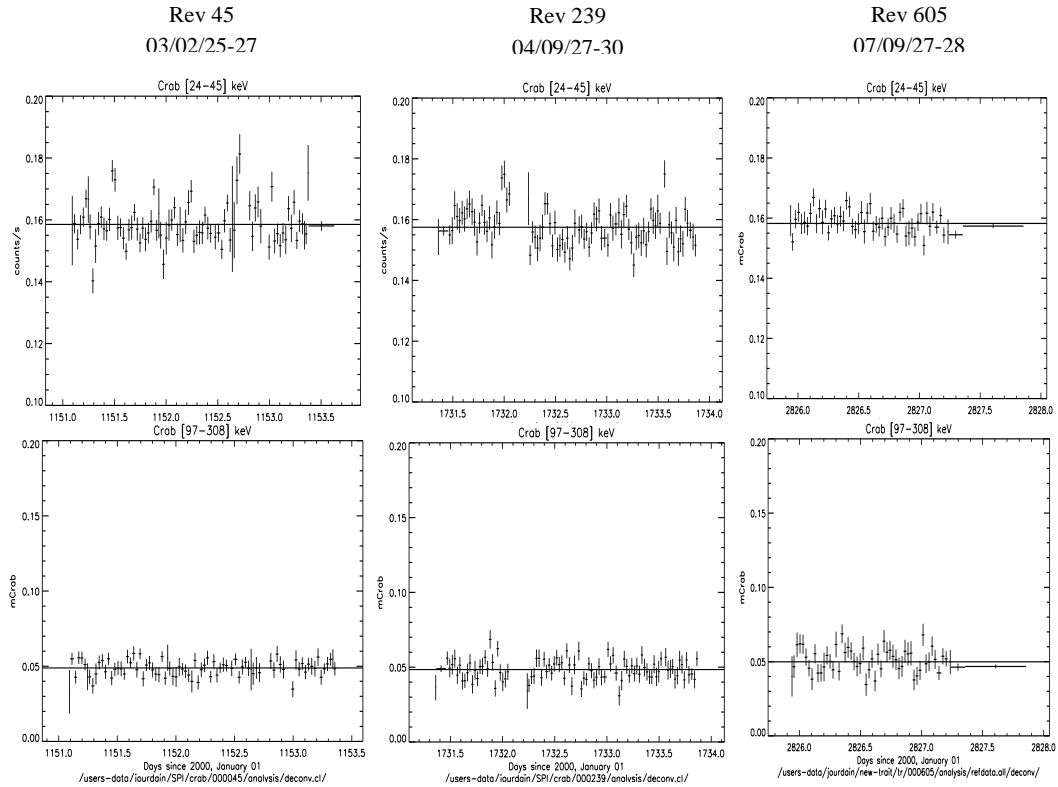

Figure 2: Light curves obtained for the Crab during three observations in two energy bands (top: 25-45 keV, bottom: 100-300 keV).
Each point corresponds to 1 Science Window (~ 2000s).

Additionnal longer observations have been required in 2008 to investigate the instrument stability up to the MeV domain.
Note that the instrument consists of only 17 detectors since the failure of 2 of them (one after revolution 140 and the second after revolution 214).

**3.2 Spectral analysis**

Based on the stability of the observed emission, we have analysed simultaneously all the observations pointed on the Crab Nebula and performed in the standard 5×5 dithering pattern (see [5] for the definition and advantages of the INTEGRAL dithering procedure), after the revolution 214 (ie in the 17 detectors configuration). We thus gathered 326 ks of data, covering 4 years of operation, from September 2004 to September 2007.

The main result, deduced from the sum (average) spectrum, is that a single power law gives a best fit photon index of 2.10 but does not adjust correctly the data as a curvature is clearly visible when covering the 20 keV to 1 MeV domain. This continuous curvature can be reasonable reproduced by two power laws, but the slopes of the power laws will depend on the position of the break. The best fit parameters can not be determined consistently from the data since they obviously vary with systematic level, energy domain or binning. We thus opt for fixing the energy break to 100 keV, to allow various comparisons between observations or with other instruments.

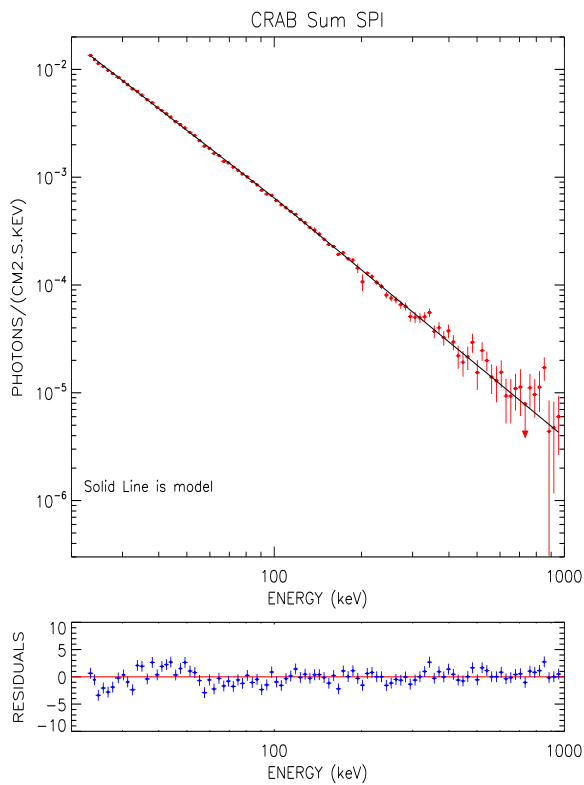

Figure 3: The SPI Crab Nebula spectrum: sum of 326 ks

*Broken power law Model*
**Best fit parameters:**
Photon Index 1    $2.08 \pm 0.02$    Break Energy    100.0 keV   (fixed)
Photon Index 2    $2.23 \pm 0.1$
1 keV Norm    $9.3 \pm 0.14$
$\chi^2 =$    143.35    (97 DoF)

Flux @ 100 keV :
$6.44 \times 10^{-4}$ ph/cm2 s keV

These parameters have been determined between 23 keV and 1 MeV with no systematics added to the data.

We have then studied the evolution of the spectral shape with time by performing similar fits for individual revolutions. We present here (see Table 1) results obtained for the same broken power law model. Fig 3 corresponds to the "Sum1" raw.

No systematics have been added and the low energy points (below 23 keV) are excluded due to instrumental/calibration problem under investigation.

All parameters vary within narrow ranges reinforcing the validity of the observed curvature in the hard X-ray domain and consequently the scientific interest to propose models able to explain this emission shape.

Table 1: Best fit parameters for individual observations

| Rev # | Duration (ks) | Index 1 | Break (keV) fixed | Index 2 | Reduced $\chi^2$ (35 dof) | Norm @ 100 keV (ph/cm2 s keV) |
|---|---|---|---|---|---|---|
| 239 | 31 | $2.07 \pm 0.02$ | 100 | $2.36 \pm 0.1$ | 1.32 | $6.35 \times 10^{-4}$ |
| 300 | 38 | $2.09 \pm 0.02$ | 100 | $2.23 \pm 0.1$ | 1.67 | $6.35 \times 10^{-4}$ |
| 365 | 30 | $2.06 \pm 0.02$ | 100 | $2.34 \pm 0.1$ | 0.99 | $6.6 \times 10^{-4}$ |
| 422 | 39 | $2.09 \pm 0.02$ | 100 | $2.20 \pm 0.1$ | 1.17 | $6.4 \times 10^{-4}$ |
| 483 | 32.5 | $2.11 \pm 0.02$ | 100 | $2.20 \pm 0.1$ | 1.85* *Non standard dithering pattern | $6.3 \times 10^{-4}$ |
| 541 | 71.5 | $2.08 \pm 0.02$ | 100 | $2.21 \pm 0.1$ | 0.95 | $6.3 \times 10^{-4}$ |
| 605 | 84 | $2.08 \pm 0.02$ | 100 | $2.20 \pm 0.1$ | 1.7 | $6.4 \times 10^{-4}$ |
| **Sum1 239-605** | **326** | **$2.08 \pm 0.01$** | **100** | **$2.23 \pm 0.05$** | **1.62 (110 dof)** | **$6.44 \times 10^{-4}$** |
| 665 | 146.5 | $2.07 \pm 0.02$ | 100 | $2.23 \pm 0.05$ | 1.23 | $6.6 \times 10^{-4}$ |
| 666 | 154 | $2.07 \pm 0.02$ | 100 | $2.23 \pm 0.05$ | 1.7 | $6.6 \times 10^{-4}$ |
| Sum2 239-666 | 626.5 | $2.07 \pm 0.02$ | 100 | $2.23 \pm 0.05$ | 2.36 (110 dof) | $6.5 \times 10^{-4}$ |

## 4. Conclusion

The stability of the SPI instrument allows us to investigate in details the high energy (20 keV -1 MeV) spectral shape of the Crab nebula. We have analyzed individual revolutions (1 per 6 months, performed for calibration purposes) and the sum of them.

Results**:**

- The spectral shape is found very stable (as well as the flux).
- The single power model is rejected by the data: A curvature is clearly required. The simplest model to adjust it is a broken power law but a continuous curvature is probably more relevant. Indeed, there is a clear degeneracy between first index and energy break, leading us to fix it to 100 keV, in order to quantify fit comparisons.

The low energy slope deduced from the SPI data (photon index of 2.08 ± 0.02) shows a good agreement with the photon indexes usually found by the X-ray instruments (2.04 from XMM [5], 2.1 from Chandra [6]) while the requirement of a break or curvature toward high energy has already been pointed out by BATSE [7] and COMPTEL measurements (photon index of 2.227 ± 0.013 in the 0.75-30 MeV domain [8]).